\documentclass[article]{jss}

\usepackage{orcidlink,thumbpdf,lmodern,amsmath}

\newcommand{\fct}[1]{\code{#1()}}

\author{Avi Kenny~\orcidlink{0000-0002-9465-7307}\\Duke University
   \And Charles J. Wolock~\orcidlink{0000-0003-3527-1102}\\University of Pennsylvania}
\Plainauthor{Avi Kenny, Charles J Wolock}

\title{SimEngine: A Modular Framework for Statistical Simulations in \proglang{R}}
\Plaintitle{SimEngine: A Modular Framework for Statistical Simulations in R}
\Shorttitle{SimEngine: A Modular Framework for Statistical Simulations in \proglang{R}}

\Abstract{
  This article describes \pkg{SimEngine}, an open-source \proglang{R} package for structuring, maintaining, running, and debugging statistical simulations on both local and cluster-based computing environments. Several \proglang{R} packages exist for structuring simulations, but \pkg{SimEngine} is the only package specifically designed for running simulations in parallel via job schedulers on high-performance cluster computing systems. The package provides structure and functionality for common simulation tasks, such as setting simulation levels, managing seeds for random number generation, and calculating summary metrics (such as bias and confidence interval coverage). \pkg{SimEngine} also brings several unique features, such as automatic calculation of Monte Carlo error and information-sharing across simulation replicates. We provide an overview of the package and demonstrate some of its advanced functionality. 
}

\Plainkeywords{simulation, cluster computing system, parallel computing, R}

\Address{
  \\Avi Kenny\\
  Department of Biostatistics and Bioinformatics\\
  Duke University School of Medicine\\
  2424 Erwin Rd\\
  Durham, NC 27710, USA\\
  E-mail: \email{avi.kenny@duke.edu}\\
  URL: \url{https://avi-kenny.github.io/}\\

  Charles J. Wolock\\
  Department of Biostatistics, Epidemiology and Informatics\\
  University of Pennsylvania\\
  423 Guardian Drive\\
  Philadelphia, PA 19104, USA\\
  E-mail: \email{cwolock@upenn.edu}\\
  URL: \url{https://cwolock.github.io/}
  
}

\begin{document}

\section[Introduction]{Introduction} \label{sec:intro}

For the past several decades, the design and execution of simulation studies has been a pillar of methodological research in statistics \citep{hauck1984survey}. Simulations are commonly used to evaluate finite sample performance of proposed statistical procedures, but can also be used to identify problems with existing methods, ensure that statistical code functions as designed, and test out new ideas in an exploratory manner. Additionally, simulation can be a statistical method in itself; two common examples are study power calculation \citep{arnold2011simulation} and sampling from a complex distribution \citep{wakefield2013bayesian}.

Although the power of personal computers has increased exponentially over the last four decades, many simulation studies require far more computing power than what is available on even a high-end laptop. Accordingly, many academic (bio)statistics departments and research institutions have invested in so-called cluster computing systems (CCS), which are essentially networks of servers available for high-throughput parallel computation. A single CCS typically serves many researchers, can be securely accessed remotely, and operates job scheduling software (e.g., Slurm) designed to coordinate the submission and management of computing tasks between multiple groups of users.

Thankfully, the majority of statistical simulations can be easily parallelized, since they typically involve running the same (or nearly the same) code many times and then performing an analysis of the results of these replicates. This allows for a simulation study to be done hundreds or thousands of times faster than if the user ran the same code on their laptop. Despite this, many potential users never leverage an available CCS; a major reason for this is that it can be difficult to get \proglang{R} and the CCS to ``work together,'' even for an experienced programmer.

To address this gap, we created \pkg{SimEngine}, an open-source \proglang{R} package \citep{Rsoftware} for structuring, maintaining, running, and debugging statistical simulations on both local and cluster-based computing environments. Several \proglang{R} packages exist for structuring simulations (e.g., \citealp{alfons2010object}, \citealp{hofert2013parallel}, \citealp{bien2016simulator}, \citealp{chalmers2020writing}, \citealp{brown2023simpr}), but \pkg{SimEngine} is the only package specifically designed for running simulations both locally and in parallel on a high-performance CCS. The package provides structure and functionality for common simulation tasks, such as setting simulation levels, managing random number generator (RNG) seeds, and calculating summary metrics (such as bias and confidence interval coverage). In addition, \pkg{SimEngine} offers a number of unique features, such as automatic calculation of Monte Carlo error \citep{koehler2009assessment} and information-sharing across simulation replicates.

This article is organized as follows. In Section \ref{sec:overview}, we give a broad overview of the \pkg{SimEngine} simulation workflow. In Section \ref{sec:parallel}, we describe how simulations can be parallelized both locally and on a CCS. Section \ref{sec:advanced} contains details on more advanced functionality of the package. The Appendix includes two example simulation studies carried out using \pkg{SimEngine}.  

\section{Overview of simulation workflow} \label{sec:overview}

The latest stable version of \pkg{SimEngine} can be installed from CRAN using \fct{install.packages}. The current development version can be installed using \fct{devtools::install\_github}.

\begin{CodeChunk}
\begin{CodeInput}
R> install.packages("SimEngine")
R> devtools::install_github(repo="Avi-Kenny/SimEngine")
\end{CodeInput}
\end{CodeChunk}

The goal of many statistical simulations is to evaluate a new statistical method against existing methods; we use this framework to demonstrate the \pkg{SimEngine} workflow. Most statistical simulations of this type include three basic phases: (1) generate data, (2) run one or more methods using the generated data, and (3) compare the performance of the methods.

To briefly illustrate how these phases are implemented using \pkg{SimEngine}, we use a simple example of estimating the rate parameter $\lambda$ of a $\text{Poisson}(\lambda)$ distribution. To anchor the simulation in a real-world situation, one can imagine that a sample of size $n$ from this Poisson distribution models the number of patients admitted to a hospital over the course of $n$ consecutive days. Suppose that the data consist of $n$ independent and identically distributed observations $X_1, X_2, \ldots, X_n$ drawn from a Poisson($\lambda$) distribution. Since the $\lambda$ parameter of the Poisson distribution is equal to both the mean and the variance, one may ask whether the sample mean $\hat{\lambda}_{M,n}:= \frac{1}{n}\sum_{i=1}^{n}X_i$ or the sample variance $\hat{\lambda}_{V,n} := \frac{1}{n-1}\sum_{i=1}^{n}(X_i - \hat{\lambda}_{M,n})^2$ is a better estimator of $\lambda$.

\subsection{Load the package and create a simulation object}

The first step is to create a simulation object (an \proglang{R} object of class \code{sim_obj}) using the \fct{new\_sim} function. The simulation object contains all data, functions, and results related to the simulation.

\begin{CodeChunk}
\begin{CodeInput}
R> library(SimEngine)
R> sim <- new_sim()
\end{CodeInput}
\end{CodeChunk}

\subsection{Code a function to generate data}

Many simulations involve one or more functions that create a dataset designed to mimic a real-world data-generating mechanism. Here, we write and test a simple function to generate a sample of \code{n} observations from a Poisson distribution with $\lambda = 20$.

\begin{CodeChunk}
\begin{CodeInput}
R> create_data <- function(n) {
+    return(rpois(n=n, lambda=20))
+  }
R> create_data(n=10)
\end{CodeInput}
\begin{CodeOutput}
 [1] 18 25 25 21 13 22 23 22 18 26
\end{CodeOutput}
\end{CodeChunk}

\subsection{Code the methods (or other functions)}

With \pkg{SimEngine}, any functions declared (or loaded via \fct{source}) are automatically added to the simulation object when the simulation runs. In this example, we test the sample mean and sample variance estimators of the $\lambda$ parameter. For simplicity, we write this as a single function and use the \code{type} argument to specify which estimator to use. One could also write two separate functions.

\begin{CodeChunk}
\begin{CodeInput}
R> est_lambda <- function(dat, type) {
+    if (type=="M") { return(mean(dat)) }
+    if (type=="V") { return(var(dat)) }
+  }
R> dat <- create_data(n=1000)
R> est_lambda(dat=dat, type="M")
\end{CodeInput}
\begin{CodeOutput}
[1] 19.646
\end{CodeOutput}
\begin{CodeInput}
R> est_lambda(dat=dat, type="V")
\end{CodeInput}
\begin{CodeOutput}
[1] 20.8195
\end{CodeOutput}
\end{CodeChunk}

\subsection{Set the simulation levels}

Often, we wish to run the same simulation multiple times. We refer to each run as a \textit{simulation replicate}. We may wish to vary certain features of the simulation between replicates. In this example, perhaps we choose to vary the sample size and the estimator used to estimate $\lambda$. We refer to the features that vary as \textit{simulation levels}; in the example below, the simulation levels are the sample size (\code{n}) and the estimator (\code{estimator}). We refer to the values that each simulation level can take on as \textit{level values}; in the example below, the \code{n} level values are \code{10}, \code{100}, and \code{1000}, and the \code{estimator} level values are \code{``M''} (for ``sample mean'') and \code{``V''} (for ``sample variance''). By default, \pkg{SimEngine} runs one simulation replicate for each combination of level value --- in this case, six combinations --- although the user will typically want to increase this; 1,000 or 10,000 replicates per combination is typical. An appropriate number of replicates per combination may be informed by the desired level of Monte Carlo error; see Section \ref{sec:mc error}.

\begin{CodeChunk}
\begin{CodeInput}
R> sim %<>% set_levels(
+    estimator = c("M", "V"),
+    n = c(10, 100, 1000)
+  )
\end{CodeInput}
\end{CodeChunk}

Note that we make extensive use of the pipe operators (\code{\%>\%} and \code{\%<>\%}) from the \pkg{magrittr} package \citep{bache2022magrittr}.

\subsection{Create a simulation script}

The simulation script is a user-written function that assembles the pieces above (generating data, analyzing the data, and returning results) to code the flow of a single simulation replicate. Within a script, the current simulation level values can be referenced using the special variable \code{L}. For instance, in the running example, when the first simulation replicate is running, \code{L\$estimator} will equal \code{``M''} and \code{L\$n} will equal \code{10}. In the next replicate, \code{L\$estimator} will equal \code{``M''} and \code{L\$n} will equal \code{100}, and so on, until all level value combinations are run. The simulation script will automatically have access to any functions that have been declared earlier in the global environment.

\begin{CodeChunk}
\begin{CodeInput}
R> sim %<>% set_script(function() {
+    dat <- create_data(n=L$n)
+    lambda_hat <- est_lambda(dat=dat, type=L$estimator)
+    return (list("lambda_hat"=lambda_hat))
+  })
\end{CodeInput}
\end{CodeChunk}

The simulation script should always return a list containing key-value pairs, where the keys are character strings. The values may be simple data types (numbers, character strings, or boolean values) or more complex data types (lists, dataframes, model objects, etc.); see Section \ref{ssec_complex}. Note that in this example, the estimators could have been coded instead as two different functions and then called from within the script using the \fct{use\_method} function; see Appendix \ref{sec:ex1} for example usage. 

\subsection{Set the simulation configuration}

The \fct{set\_config} function controls options related to the entire simulation, such as the number of simulation replicates to run for each level value combination and the parallelization type, if desired (see Section \ref{sec:parallel}). Packages needed for the simulation should be specified using the \code{packages} argument of \fct{set\_config} (rather than using \fct{library} or \fct{require}). We set \code{num_sim} to 100, and so \pkg{SimEngine} will run a total of 600 simulation replicates (100 for each of the six level value combinations).

\begin{CodeChunk}
\begin{CodeInput}
R> sim %<>% set_config(
+    num_sim = 100,
+    packages = c("ggplot2", "stringr")
+  )
\end{CodeInput}
\end{CodeChunk}

\subsection{Run the simulation}

All 600 replicates are run at once and results are stored in the simulation object.

\begin{CodeChunk}
\begin{CodeInput}
R> sim %<>% run()
\end{CodeInput}
\begin{CodeOutput}
  |########################################| 100%
Done. No errors or warnings detected.
\end{CodeOutput}
\end{CodeChunk}

\subsection{View and summarize results}

Once the simulations have finished, the \fct{summarize} function can be used to calculate common summary statistics, such as bias, variance, mean squared error (MSE), and confidence interval coverage.

\begin{CodeChunk}
\begin{CodeInput}
R> sim %>% summarize(
+    list(stat="bias", name="bias_lambda", estimate="lambda_hat", truth=20),
+    list(stat="mse", name="mse_lambda", estimate="lambda_hat", truth=20)
+  )
\end{CodeInput}
\begin{CodeOutput}
  level_id estimator    n n_reps bias_lambda   mse_lambda
1        1         M   10    100 -0.17600000   1.52480000
2        2         V   10    100 -1.80166667 103.23599630
3        3         M  100    100 -0.03770000   0.19165100
4        4         V  100    100  0.22910707   7.72262714
5        5         M 1000    100  0.01285000   0.01553731
6        6         V 1000    100  0.02514744   0.92133037
\end{CodeOutput}
\end{CodeChunk}

In this example, we see that the MSE of the sample variance is much higher than that of the sample mean and that MSE decreases with increasing sample size for both estimators, as expected. From the \code{n_reps} column, we see that 100 replicates were successfully run for each level value combination. Results for individual simulation replicates can also be directly accessed via the \code{sim$results} dataframe.

\begin{CodeChunk}
\begin{CodeInput}
R> head(sim$results)
\end{CodeInput}
\begin{CodeOutput}
  sim_uid level_id rep_id estimator  n      runtime lambda_hat
1       1        1      1         M 10 0.0003290176       20.1
2       7        1      2         M 10 0.0002038479       20.5
3       8        1      3         M 10 0.0001709461       17.3
4       9        1      4         M 10 0.0001630783       20.3
5      10        1      5         M 10 0.0001599789       18.3
6      11        1      6         M 10 0.0001561642       20.4
\end{CodeOutput}
\end{CodeChunk}

Above, the \code{sim_uid} uniquely identifies a single simulation replicate and the \code{level_id} uniquely identifies a level value combination. The \code{rep_id} is unique within a given level value combination and identifies the index of that replicate within the level value combination.

\subsection{Update a simulation}\label{ssec_update}

After running a simulation, a user may want to update it by adding additional levels or replicates; this can be done with the \fct{update\_sim} function. Prior to running \fct{update\_sim}, the functions \fct{set\_levels} and/or \fct{set\_config} are used to declare the updates that should be performed. For example, the following code sets the total number of replicates to 200 (i.e., adding 100 replicates to those that have already been run) for each level value combination, and adds one additional level value for \code{n}.

\begin{CodeChunk}
\begin{CodeInput}
R> sim %<>% set_config(num_sim = 200)
R> sim %<>% set_levels(
+    estimator = c("M", "V"),
+    n = c(10, 100, 1000, 10000)
+  )
\end{CodeInput}
\end{CodeChunk}

After the levels and/or configuration are updated, \fct{update\_sim} is called.

\begin{CodeChunk}
\begin{CodeInput}
R> sim %<>% update_sim()
\end{CodeInput}
\begin{CodeOutput}
  |########################################| 100%
Done. No errors or warnings detected.
\end{CodeOutput}
\end{CodeChunk}

Another call to \fct{summarize} shows that the additional replicates were successfully:

\begin{CodeChunk}
\begin{CodeInput}
R> sim %>% summarize(
+    list(stat="bias", name="bias_lambda", estimate="lambda_hat", truth=20),
+    list(stat="mse", name="mse_lambda", estimate="lambda_hat", truth=20)
+  )
\end{CodeInput}
\begin{CodeOutput}
  level_id estimator     n n_reps  bias_lambda   mse_lambda
1        1         M    10    200 -0.205500000  1.875450000
2        2         V    10    200 -1.189166667 96.913110494
3        3         M   100    200 -0.055000000  0.197541000
4        4         V   100    200  0.023244949  7.955606709
5        5         M  1000    200  0.017495000  0.017497115
6        6         V  1000    200  0.053941807  0.874700025
7        7         M 10000    200 -0.005233000  0.002096102
8        8         V 10000    200 -0.007580998  0.072997135
\end{CodeOutput}
\end{CodeChunk}

\section{Parallelization}\label{sec:parallel}

User-friendly parallelization is a hallmark of \pkg{SimEngine}. There are two modes of parallelizing code using \pkg{SimEngine}, which we refer to as \textit{local parallelization} and \textit{cluster parallelization}.  Most statistical simulations involve running multiple replicates of the same experiment, perhaps with certain factors changing between replicates. With local parallelization, individual simulation replicates are run as single tasks on the separate cores of a local machine. Cluster parallelization occurs on a CCS. Each simulation replicate is assigned to a single task, and tasks are submitted as a job array to the CCS. \pkg{SimEngine} is designed to automate as much of the parallelization process as possible. We give an overview of each parallelization mode below. 

\subsection{Local parallelization}
Local parallelization is the most straightforward way to parallelize code, as the entire process is handled by the package and executed locally. This mode is implemented by specifying \code{parallel=TRUE} using \fct{set\_config}. \pkg{SimEngine} handles the actual parallelization task internally using the base \proglang{R} package \pkg{parallel} \citep{Rsoftware}. 

\begin{CodeChunk}
\begin{CodeInput}  
sim <- new_sim()
sim %<>% set_config(parallel = TRUE)
\end{CodeInput}
\end{CodeChunk}

If a single simulation replicate runs in a very short amount of time (e.g., less than one second), using local parallelization can actually result in an \textit{increase} in total runtime. This is because there is a certain amount of computational overhead involved in setting up the parallelization engine inside \pkg{SimEngine}. A speed comparison can be performed by running the code twice, once with \code{set\_config(parallel = TRUE)} and once with \code{set\_config(parallel = FALSE)}, each followed by \code{sim \%>\% vars("total_runtime")} to see the difference in total runtime. The exact overhead involved with local parallelization will differ between machines.

If a machine has \code{n} cores available, \pkg{SimEngine} will use \code{n-1} cores by default. The \code{n\_cores} option of the \fct{set\_config} function can be used to manually specify the number of cores to use, e.g., \code{sim \%<>\% set_config(n_cores = 2)}.

\subsection{Cluster parallelization}

Parallelizing code using a CCS is more complicated, but \pkg{SimEngine} is built to streamline this process as much as possible. A CCS is a supercomputer that consists of a number of nodes, each of which may have multiple cores. Typically, a user logs into the CCS and runs programs by submitting ``jobs'' to the CCS using a special program called a job scheduler. The job scheduler manages the process of running the jobs in parallel across multiple nodes and/or multiple cores.

Although there are multiple ways to run code in parallel on a CCS, \pkg{SimEngine} makes use of job arrays. The main cluster parallelization function in \pkg{SimEngine} is \fct{run\_on\_cluster}. Throughout this example, we use Slurm as an example job scheduler, but an analogous workflow will apply to other job scheduler software. Suppose we have written the following simulation and wish to run it on a CCS:

\begin{CodeChunk}
\begin{CodeInput}
sim <- new_sim()
create_data <- function(n) { return(rpois(n=n, lambda=20)) }
est_lambda <- function(dat, type) {
  if (type=="M") { return(mean(dat)) }
  if (type=="V") { return(var(dat)) }
}
sim %<>% set_levels(estimator = c("M","V"), n = c(10,100,1000))
sim %<>% set_script(function() {
  dat <- create_data(L$n)
  lambda_hat <- est_lambda(dat=dat, type=L$estimator)
  return(list("lambda_hat"=lambda_hat))
})
sim %<>% set_config(num_sim=100)
sim %<>% run()
sim %>% summarize()
\end{CodeInput}
\end{CodeChunk}

To run this code on a CCS, we wrap it in the \fct{run\_on\_cluster} function. To use this function, we must break the code into three blocks, called \code{first}, \code{main}, and \code{last}. The code in the \code{first} block will run only once, and will set up the simulation object. When this is finished, \pkg{SimEngine} will save the simulation object in the file system of the CCS. The code in the \code{main} block will then run once for each simulation replicate, and will have access to the simulation object created in the \code{first} block. In most cases, the code in the \code{main} block will simply include a single call to \fct{run} (or to \fct{update\_sim}, as detailed in Section \ref{sssec_update_sim_on_cluster}). Finally, the code in the \code{last} block will run after all simulation replicates have finished running, and after \pkg{SimEngine} has automatically compiled the results into the simulation object. Use of the \fct{run\_on\_cluster} function is illustrated below:

\begin{CodeChunk}
\begin{CodeInput}
run_on_cluster(
  first = {
    sim <- new_sim()
    create_data <- function(n) { return(rpois(n=n, lambda=20)) }
    est_lambda <- function(dat, type) {
      if (type=="M") { return(mean(dat)) }
      if (type=="V") { return(var(dat)) }
    }
    sim %<>% set_levels(estimator = c("M","V"), n = c(10,100,1000))
    sim %<>% set_script(function() {
      dat <- create_data(L$n)
      lambda_hat <- est_lambda(dat=dat, type=L$estimator)
      return(list("lambda_hat"=lambda_hat))
    })
    sim %<>% set_config(num_sim=100)
  },
  main = {
    sim %<>% run()
  },
  last = {
    sim %>% summarize()
  },
  cluster_config = list(js="slurm")
)        
\end{CodeInput}
\end{CodeChunk}

Note that none of the actual simulation code changed from the previous code block; we simply divided the code into chunks and and placed these chunks into the appropriate \code{first}, \code{main}, or \code{last} block within \fct{run\_on\_cluster}. Additionally, we specified  which job scheduler to use in the \code{cluster\_config} argument list. The command \fct{js\_support} can be run in \proglang{R} to see a list of supported job scheduler software; the value in the \code{js\_code} column is the value that should be specified in the \code{cluster\_config} argument. Unsupported job schedulers can still be used for cluster parallelization, as detailed below.

Next, we must give the job scheduler instructions on how to run the above code. In the following, we assume that the \proglang{R} code above is stored in a file called \code{my\_simulation.R} on the CCS file system. We create a simple shell script called \code{run\_sim.sh} with the following two lines, which will run \code{my\_simulation.R}, and place it in the same directory as the \code{my\_simulation.R} file. We demonstrate this using \proglang{BASH} scripting language, but any shell scripting language may be used.

\begin{CodeChunk}
\begin{CodeInput}
> #!/bin/bash
> Rscript my_simulation.R
\end{CodeInput}
\end{CodeChunk}

Finally, we use the job scheduler to submit three jobs. The first will run the \code{first} code, the second will run the \code{main} code, and the third will run the \code{last} code. With Slurm, we run the following three shell commands:

\begin{CodeChunk}
\begin{CodeInput}
> sbatch --export=sim_run='first' run_sim.sh
\end{CodeInput}
\begin{CodeOutput}
Submitted batch job 101
\end{CodeOutput}
\begin{CodeInput}
> sbatch --export=sim_run='main' --array=1-20 --depend=afterok:101 run_sim.sh
\end{CodeInput}
\begin{CodeOutput}
Submitted batch job 102
\end{CodeOutput}
\begin{CodeInput}
> sbatch --export=sim_run='last' --depend=afterok:102 run_sim.sh
\end{CodeInput}
\begin{CodeOutput}
Submitted batch job 103
\end{CodeOutput}
\end{CodeChunk}

In the first line, we submit the \code{run\_sim.sh} script using the \code{sim\_run=`first'} environment variable, which tells \pkg{SimEngine} to only run the code in the \code{first} block within the \fct{run\_on\_cluster} function in \code{my\_simulation.R}. After running this, Slurm returns the message \code{Submitted batch job 101}. The number \code{101} is called the ``job ID'' and uniquely identifies the job on the CCS.

In the second line, we submit the \code{run\_sim.sh} script using the \code{sim\_run=`main'} environment variable and tell Slurm to run a job array with ``task IDs'' 1-20. Each task will occupy one core, so, in this case, 20 cores will be used. The user is responsible for specifying the number of cores to use. By default, each simulation replicate is assigned to run on its own core, and so in this case, the number of cores should be set to correspond with the total number of replicates in the simulation. However, it is also possible to run multiple replicates per core; see below. 
Also note that we included the option \code{--depend=afterok:101}, which instructs the job scheduler to wait until the first job finishes before starting the job array. (In practice, the number 101 must be replaced with whatever job ID Slurm assigned to the first job.) Once this command is submitted, the code in the \code{main} block will be run for each replicate. A temporary folder called \code{sim\_results} will be created and filled with objects representing the results and/or errors for each replicate.

In the third line, we submit the \code{run\_sim.sh} script using the \code{sim\_run='last'} environment variable. Again, we use \code{--depend=afterok:102} to ensure this code does not run until all tasks in the job array have finished. When this job runs, \pkg{SimEngine} will compile the results from the \code{main} block, run the code in the \code{last} block, save the simulation object to the file system, and delete the temporary \code{sim\_results} folder and its contents.

As mentioned above, the default \pkg{SimEngine} behavior is to run each replicate on its own core. However, it can sometimes be advantageous to run multiple replicates per core (e.g., if a single replicate has a short runtime, or if the job scheduler limits how many job array tasks can be submitted at one time). To do this, we first specify the \code{n\_cores} option using \fct{set\_config}, e.g., \code{sim \%<>\% set_config(n_cores = 5)}. Second, we change the second \code{sbatch} command to match the value used for \code{n\_cores}:

\begin{CodeChunk}
\begin{CodeInput}
> sbatch --export=sim_run='main' --array=1-5 --depend=afterok:101 run_sim.sh
\end{CodeInput}
\begin{CodeOutput}
Submitted batch job 102
\end{CodeOutput}
\end{CodeChunk}

Here, only five cores will be used to run the 20 simulation replicates, and so only five job array tasks are needed.

\subsection{Additional cluster parallelization functionality}

\subsubsection{Running locally}

The \fct{run\_on\_cluster} function is programmed such that it can also be run locally. In this case, the code within the \code{first}, \code{main}, and \code{last} blocks will be executed on the local machine rather than on a CCS. Objects created within these three blocks will not be saved, but a copy of the simulation object will be exported to the environment from which \fct{run\_on\_cluster} was called (typically the global environment) so that the results may be examined, passed to functions like \fct{summarize}, etc. This can be useful for testing simulations locally before sending them to a CCS.

\subsubsection{Using unsupported job schedulers}

There may be job schedulers that \pkg{SimEngine} does not natively support. If this is the case, \pkg{SimEngine} can still be used for cluster parallelization; the key is to identify the environment variable that the job scheduler uses to uniquely identify tasks within a job array. For example, Slurm uses the variable \code{"SLURM_ARRAY_TASK_ID"} and Grid Engine uses the variable \code{"SGE_TASK_ID"}. Once this variable is identified, it can be specified in the \code{cluster\_config} block, as follows:

\begin{CodeChunk}
\begin{CodeInput}
run_on_cluster(
  first = {...},
  main = {...},
  last = {...},
  cluster_config = list(tid_var="SLURM_ARRAY_TASK_ID")
)
\end{CodeInput}
\end{CodeChunk}

\subsubsection{Updating a simulation on a CCS}\label{sssec_update_sim_on_cluster}

To update a simulation on a CCS, the \fct{update\_sim\_on\_cluster} function can be used. The workflow is similar to that of \fct{run\_on\_cluster}, with several key differences. Instead of creating a new simulation object in the \code{first} block using \fct{new\_sim}, the existing simulation object (which would have been saved to the filesystem when \fct{run\_on\_cluster} was called originally) is loaded using \fct{readRDS}. Then, the functions \fct{set\_levels} and/or \fct{set\_config} are called to specify the desired updates (see Section \ref{ssec_update}). In the \code{main} block, \fct{update\_sim} is called (instead of \fct{run}). In the \code{last} block, code can remain the same or change as needed. These differences are illustrated in the code below.

\begin{CodeChunk}
\begin{CodeInput}
update_sim_on_cluster(
  first = {
    sim <- readRDS("sim.rds")
    sim %<>% set_levels(n=c(100,500,1000))
  },
  main = {
    sim %<>% update_sim()
  },
  last = {
    sim %>% summarize()
  },
  cluster_config = list(js="slurm")
)        
\end{CodeInput}
\end{CodeChunk}

Submission of this code via a job scheduler proceeds in the same manner as described earlier for \fct{run\_on\_cluster}.

\section{Advanced functionality}\label{sec:advanced}

In this section, we review the following functionality, targeting advanced users of the package:

\begin{itemize}
  \item Using the powerful \fct{batch} function, which allows for information to be shared across simulation replicates;
  \item Using \textit{complex simulation levels} in settings where simple levels (e.g., a vector of numbers) are insufficient;
  \item Handling \textit{complex return data} in settings where a single simulation replicate returns nested lists, dataframes, model objects, etc.;
  \item Managing RNG seeds;
  \item Best practices for debugging and handling errors and warnings;
  \item Capturing Monte Carlo error using the \fct{summarize} function. 
\end{itemize}

\subsection{Using the batch function to share information across simulation replicates}

The \fct{batch} function is useful for sharing data or objects between simulation replicates. Essentially, it allows simulation replicates to be divided into ``batches;'' all replicates in a given batch will then share a single set of objects. A common use case for this is a simulation that involves using multiple methods to analyze a shared dataset, and repeating this process over a number of dataset replicates. This may be of interest if, for example, it is computationally expensive to generate a simulated dataset.  

To illustrate the use of \fct{batch} using this example, we first consider the following simulation:

\begin{CodeChunk}
\begin{CodeInput}
R> sim <- new_sim()
R> create_data <- function(n) { rnorm(n=n, mean=3) }
R> est_mean <- function(dat, type) {
+    if (type=="est_mean") { return(mean(dat)) }
+    if (type=="est_median") { return(median(dat)) }
+  }
R> sim %<>% set_levels(est=c("est_mean","est_median"))
R> sim %<>% set_config(num_sim=3)
R> sim %<>% set_script(function() {
+    dat <- create_data(n=100)
+    mu_hat <- est_mean(dat=dat, type=L$est)
+    return(list(
+      "mu_hat" = round(mu_hat,2),
+      "dat_1" = round(dat[1],2)
+    ))
+  })
R> sim %<>% run()
\end{CodeInput}
\begin{CodeOutput}
  |########################################| 100%
Done. No errors or warnings detected.
\end{CodeOutput}
\end{CodeChunk}

From the \code{``dat\_1''} column of the results object (equal to the first element of the \code{dat} vector created in the simulation script), we see that a unique dataset was created for each simulation replicate:

\begin{CodeChunk}
\begin{CodeInput}
R> sim$results[order(sim$results$rep_id),c(1:7)!=5]
\end{CodeInput}
\begin{CodeOutput}
  sim_uid level_id rep_id        est mu_hat dat_1
1       1        1      1   est_mean   3.05  4.09
4       2        2      1 est_median   3.06  3.23
2       3        1      2   est_mean   3.03  2.99
5       5        2      2 est_median   3.02  2.78
3       4        1      3   est_mean   2.85  1.47
6       6        2      3 est_median   3.03  2.35
\end{CodeOutput}
\end{CodeChunk}

Suppose that instead, we wish to analyze each simulated dataset using multiple methods (in this case corresponding to \code{``est\_mean''} and \code{``est\_median''}), and repeat this procedure a total of three times. We can do this using the \fct{batch} function, as follows:

\begin{CodeChunk}
\begin{CodeInput}
R> sim <- new_sim()
R> create_data <- function(n) { rnorm(n=n, mean=3) }
R> est_mean <- function(dat, type) {
+    if (type=="est_mean") { return(mean(dat)) }
+    if (type=="est_median") { return(median(dat)) }
+  }
R> sim %<>% set_levels(est=c("est_mean","est_median"))
R> sim %<>% set_config(num_sim=3, batch_levels=NULL)
R> sim %<>% set_script(function() {
+    batch({
+      dat <- create_data(n=100)
+    })
+    mu_hat <- est_mean(dat=dat, type=L$est)
+    return(list(
+      "mu_hat" = round(mu_hat,2),
+      "dat_1" = round(dat[1],2)
+    ))
+  })
R> sim %<>% run()
\end{CodeInput}
\begin{CodeOutput}
  |########################################| 100%
Done. No errors or warnings detected.
\end{CodeOutput}
\end{CodeChunk}

In the code above, we changed two things. First, we added \code{batch_levels=NULL} to the \code{set_config} call; this will be explained below. Second, we wrapped the code line \code{dat <- create_data(n=100)} inside the \fct{batch} function. Whatever code goes inside the \fct{batch} function will produce the same output for all simulations in a batch; in this case, from the \code{``dat_1''} column of the results object, we see that one dataset was created and shared by the batch corresponding to \code{sim\_uids} 1 and 2 (likewise for \code{sim\_uids} $\{3,4\}$ and $\{5,6\}$):

\begin{CodeChunk}
\begin{CodeInput}
R> sim$results[order(sim$results$rep_id),c(1:7)!=5]
\end{CodeInput}
\begin{CodeOutput}
  sim_uid level_id rep_id        est mu_hat dat_1
1       1        1      1   est_mean   3.02  2.74
4       2        2      1 est_median   3.19  2.74
2       3        1      2   est_mean   2.91  3.71
5       5        2      2 est_median   2.95  3.71
3       4        1      3   est_mean   3.10  3.52
6       6        2      3 est_median   3.01  3.52
\end{CodeOutput}
\end{CodeChunk}

However, the situation is often more complicated. Suppose we have a simulation with multiple level variables, some that correspond to creating data and some that correspond to analyzing the data. Here, the \code{batch_levels} config option plays a role. We simply specify the names of the level variables that are used within the \fct{batch} function. In the example below, the levels \code{n} and \code{mu} are used within batches, while the level \code{est} is not. 

\begin{CodeChunk}
\begin{CodeInput}
R> sim <- new_sim()
R> create_data <- function(n, mu) { rnorm(n=n, mean=mu) }
R> est_mean <- function(dat, type) {
+    if (type=="est_mean") { return(mean(dat)) }
+    if (type=="est_median") { return(median(dat)) }
+  }
R> sim %<>% set_levels(n=c(10,100), mu=c(3,5), est=c("est_mean","est_median"))
R> sim %<>% set_config(num_sim=2, batch_levels=c("n", "mu"), return_batch_id=T)
R> sim %<>% set_script(function() {
+    batch({
+      dat <- create_data(n=L$n, mu=L$mu)
+    })
+    mu_hat <- est_mean(dat=dat, type=L$est)
+    return(list(
+      "mu_hat" = round(mu_hat,2),
+      "dat_1" = round(dat[1],2)
+    ))
+  })
R> sim %<>% run()
\end{CodeInput}
\begin{CodeOutput}
  |########################################| 100%
Done. No errors or warnings detected.
\end{CodeOutput}
\begin{CodeInput}
R> sim$results[order(sim$results$batch_id),c(1:10)!=8]
\end{CodeInput}
\begin{CodeOutput}
   sim_uid level_id rep_id batch_id   n mu        est mu_hat dat_1
1        1        1      1        1  10  3   est_mean   2.87  4.29
9        5        5      1        1  10  3 est_median   2.94  4.29
2        9        1      2        2  10  3   est_mean   2.79  2.77
10      13        5      2        2  10  3 est_median   2.73  2.77
3        2        2      1        3 100  3   est_mean   2.93  1.77
11       6        6      1        3 100  3 est_median   3.01  1.77
4       10        2      2        4 100  3   est_mean   2.80  4.44
12      14        6      2        4 100  3 est_median   2.71  4.44
5        3        3      1        5  10  5   est_mean   5.49  4.78
13       7        7      1        5  10  5 est_median   5.25  4.78
6       11        3      2        6  10  5   est_mean   4.57  4.48
14      15        7      2        6  10  5 est_median   4.62  4.48
7        4        4      1        7 100  5   est_mean   4.98  5.66
15       8        8      1        7 100  5 est_median   4.95  5.66
8       12        4      2        8 100  5   est_mean   5.08  5.55
16      16        8      2        8 100  5 est_median   5.14  5.55
\end{CodeOutput}
\end{CodeChunk}

The batches were created such that each batch contained two replicates, one for each \code{est}. For expository purposes, we also specified the \code{return_batch_id=T} option in \code{set_config} so that the results object would return the \code{batch_id}. This is not necessary in practice. The \code{batch_id} variable defines the batches; all simulations that share the same \code{batch_id} are in a single batch. The \code{return_batch_id=T} option can be useful to ensure correct usage of the \fct{batch} function.

We note the following about the \fct{batch} function:

\begin{itemize}
  \item The code within the \fct{batch} code block must \textit{only} create objects; this code should not change or delete existing objects, as these changes will be ignored.
  \item In the majority of cases, the \fct{batch} function will be called just once, at the beginning of the simulation script. However, it can be used anywhere in the script and can be called multiple times. The \fct{batch} function should never be used outside of the simulation script.
  \item Although we have illustrated the use of the \fct{batch} function to create a dataset to share between multiple simulation replicates, it can be used for much more, e.g., taking a sample from an existing dataset, computing shared nuisance function estimators, performing minimzing the number of repetitions of a computationally intensive task, etc.
  \item If the simulation is being run in parallel (either locally or on a CCS), \code{n_cores} must be set, and the number of cores used cannot exceed the number of batches, since all simulations within a batch must run on the same core.
  \item If the simulation script uses the \fct{batch} function, the simulation cannot be updated using the \fct{update\_sim} or \fct{update\_sim\_on\_cluster} functions, with the exception of updates that only entail removing simulation replicates.
\end{itemize}


\subsection{Complex simulation levels}

Often, simulation levels are simple, such as a vector of sample sizes:

\begin{CodeChunk}
\begin{CodeInput}
R> sim <- new_sim()
R> sim %<>% set_levels(n = c(200,400,800))
\end{CodeInput}
\end{CodeChunk}

However, there are many instances in which more complex objects are needed. For these cases, instead of a vector of numbers or character strings, we use a named list of lists. The toy example below illustrates this.

\begin{CodeChunk}
\begin{CodeInput}
R> sim <- new_sim()
R> sim %<>% set_levels(
+    n = c(10,100),
+    distribution = list(
+      "Beta 1" = list(type="Beta", params=c(0.3, 0.7)),
+      "Beta 2" = list(type="Beta", params=c(1.5, 0.4)),
+      "Normal" = list(type="Normal", params=c(3.0, 0.2))
+    )
+  )
R> create_data <- function(n, type, params) {
+    if (type=="Beta") {
+      return(rbeta(n, shape1=params[1], shape2=params[2]))
+    } else if (type=="Normal") {
+      return(rnorm(n, mean=params[1], sd=params[2]))
+    }
+  }
R> sim %<>% set_script(function() {
+    x <- create_data(L$n, L$distribution$type, L$distribution$params)
+    return(list("y"=mean(x)))
+  })
R> sim %<>% run()
\end{CodeInput}
\begin{CodeOutput}
  |########################################| 100%
Done. No errors or warnings detected.
\end{CodeOutput}
\end{CodeChunk}

Note that the list names (\code{``Beta 1''}, \code{``Beta 2''}, and \code{``Normal''}) become the entries in the \code{sim\$results} dataframe, as well as the dataframe returned by \fct{summarize}.

\begin{CodeChunk}
\begin{CodeInput}
R> sim %>% summarize(list(stat="mean", x="y"))
\end{CodeInput}
\begin{CodeOutput}
  level_id   n distribution n_reps    mean_y
1        1  10       Beta 1      1 0.1635174
2        2 100       Beta 1      1 0.2740965
3        3  10       Beta 2      1 0.6234866
4        4 100       Beta 2      1 0.7664522
5        5  10       Normal      1 3.0903062
6        6 100       Normal      1 3.0179944
\end{CodeOutput}
\end{CodeChunk}

\subsection{Complex return data}\label{ssec_complex}

In most situations, the results of simulations are numeric. However, we may want to return more complex data, such as matrices, lists, or model objects. To do this, we add a key-value pair to the list returned by the simulation script with the special key \code{".complex"} and a list (containing the complex data) as the value. This is illustrated in the toy example below. Here, the simulation script estimates the parameters of a linear regression and returns these as numeric, but also returns the estimated covariance matrix and the entire model object.

\begin{CodeChunk}
\begin{CodeInput}
R> sim <- new_sim()
R> sim %<>% set_levels(n=c(10, 100, 1000))
R> create_data <- function(n) {
+    x <- runif(n)
+    y <- 3 + 2*x + rnorm(n)
+    return(data.frame("x"=x, "y"=y))
+  }
R> sim %<>% set_config(num_sim=2)
R> sim %<>% set_script(function() {
+    dat <- create_data(L$n)
+    model <- lm(y~x, data=dat)
+    return(list(
+      "beta0_hat" = model$coefficients[[1]],
+      "beta1_hat" = model$coefficients[[2]],
+      ".complex" = list(
+        "model" = model,
+        "cov_mtx" = vcov(model)
+      )
+    ))
+  })
R> sim %<>% run()
\end{CodeInput}
\begin{CodeOutput}
  |########################################| 100%
Done. No errors or warnings detected.
\end{CodeOutput}
\end{CodeChunk}

After running this simulation, we can examine the numeric results directly by accessing \code{sim\$results} or using the \code{summarize} function, as usual:

\begin{CodeChunk}
\begin{CodeInput}
R> head(sim$results)
\end{CodeInput}
\begin{CodeOutput}
  sim_uid level_id rep_id    n     runtime beta0_hat beta1_hat
1       1        1      1   10 0.012123823  2.113012  3.780972
2       4        1      2   10 0.001345873  2.058610  3.726216
3       2        2      1  100 0.006520033  2.784147  2.058041
4       5        2      2  100 0.001568794  3.066045  2.097569
5       3        3      1 1000 0.001888037  3.144964  1.783498
6       6        3      2 1000 0.002427101  3.125128  1.714405
\end{CodeOutput}
\end{CodeChunk}

However, we may also want to look at the complex return data. To do so, we use the special function \code{get_complex}, as illustrated below:

\begin{CodeChunk}
\begin{CodeInput}
R> c5 <- get_complex(sim, sim_uid=5)
R> print(summary(c5$model))
\end{CodeInput}
\begin{CodeOutput}
Call:
lm(formula = y ~ x, data = dat)

Residuals:
    Min      1Q  Median      3Q     Max 
-2.7050 -0.7429  0.1183  0.7470  1.9673 

Coefficients:
            Estimate Std. Error t value Pr(>|t|)    
(Intercept)   3.0660     0.2127  14.413  < 2e-16 ***
x             2.0976     0.3714   5.647 1.59e-07 ***
---
Signif. codes:  0 ‘***’ 0.001 ‘**’ 0.01 ‘*’ 0.05 ‘.’ 0.1 ‘ ’ 1

Residual standard error: 1.003 on 98 degrees of freedom
Multiple R-squared:  0.2455,	Adjusted R-squared:  0.2378 
F-statistic: 31.89 on 1 and 98 DF,  p-value: 1.593e-07
\end{CodeOutput}
\begin{CodeInput}
R> print(c5$cov_mtx)
\end{CodeInput}
\begin{CodeOutput}
            (Intercept)           x
(Intercept)  0.04525148 -0.06968649
x           -0.06968649  0.13795995
\end{CodeOutput}
\end{CodeChunk}

\subsection{Random number generator seeds}

In statistical research, it is typically desirable to be able to reproduce the exact results of a simulation study. Since \proglang{R} code often involves stochastic (random) functions like \fct{rnorm} or \fct{sample} that return different values when called multiple times, reproducibility is not guaranteed. In a simple \proglang{R} script, calling the \fct{set.seed} function at the beginning of the script ensures that the code that follows will produce the same results whenever the script is run. However, a more nuanced strategy is needed when running simulations. When running 100 replicates of the same simulation, we do not want each replicate to return identical results; rather, we would like for each replicate to be different from one another, but for \textit{the entire set of replicates} to be the same when the entire simulation is run twice in a row. \pkg{SimEngine} manages this process, even when simulations are being run in parallel locally or on a cluster computing system. \pkg{SimEngine} uses a single ``global seed'' that changes the individual seeds for each simulation replicate. The \fct{set\_config} function is used to set or change this global seed:

\begin{CodeChunk}
\begin{CodeInput}
R> sim %<>% set_config(seed=123)
\end{CodeInput}
\end{CodeChunk}

If a seed is not set using \fct{set\_config}, \pkg{SimEngine} will set a random seed automatically so that the results can be replicated if desired. To view this seed, we use the \fct{vars} function:

\begin{CodeChunk}
\begin{CodeInput}
R> sim <- new_sim()
R> print(vars(sim, "seed"))
\end{CodeInput}
\begin{CodeOutput}
[1] 287577520
\end{CodeOutput}
\end{CodeChunk}

\subsection{Debugging and error/warning handling}

In the simulation coding workflow, errors are inevitable. Some errors may affect all simulation replicates, while other errors may only affect a subset of replicates. By default, when a simulation is run, \pkg{SimEngine} will not stop if an error occurs; instead, errors are logged and stored in a dataframe along with information about the simulation replicates that resulted in those errors. Examining this dataframe by typing \code{print(sim\$errors)} can sometimes help to quickly pinpoint the issue. This is demonstrated below:

\begin{CodeChunk}
\begin{CodeInput}
R> sim <- new_sim()
R> sim %<>% set_config(num_sim=2)
R> sim %<>% set_levels(
+    Sigma = list(
+      s1 = list(mtx=matrix(c(3,1,1,2), nrow=2)),
+      s3 = list(mtx=matrix(c(4,3,3,9), nrow=2)),
+      s2 = list(mtx=matrix(c(1,2,2,1), nrow=2)),
+      s4 = list(mtx=matrix(c(8,2,2,6), nrow=2))
+    )
+  )
R> sim %<>% set_script(function() {
+    x <- MASS::mvrnorm(n=1, mu=c(0,0), Sigma=L$Sigma$mtx)
+    return(list(x1=x[1], x2=x[2]))
+  })
R> sim %<>% run()
\end{CodeInput}
\begin{CodeOutput}
  |########################################| 100%
Done. Errors detected in 25% of simulation replicates. Warnings detected in
0% of simulation replicates.
\end{CodeOutput}
\begin{CodeInput}
R> print(sim$errors)
\end{CodeInput}
\begin{CodeOutput}
  sim_uid level_id rep_id Sigma      runtime                          message
1       5        3      1    s2 0.0004692078 'Sigma' is not positive definite
2       6        3      2    s2 0.0006608963 'Sigma' is not positive definite
                                                     call
1 MASS::mvrnorm(n = 1, mu = c(0, 0), Sigma = L$Sigma$mtx)
2 MASS::mvrnorm(n = 1, mu = c(0, 0), Sigma = L$Sigma$mtx)
\end{CodeOutput}
\end{CodeChunk}

From the output above, we see that the code fails for the simulation replicates that use the level with \code{Sigma="s2"} because it uses an invalid (not positive definite) covariance matrix. Similarly, if a simulation involves replicates that throw warnings, all warnings are logged and stored in the dataframe \code{sim\$warnings}.

The workflow demonstrated above can be useful to quickly spot errors, but it has two main drawbacks. First, it is undesirable to run a time-consuming simulation involving hundreds or thousands of replicates, only to find at the end that every replicate failed because of a typo. It may therefore useful to stop an entire simulation after a single error has occurred. Second, it can sometimes be difficult to determine exactly what caused an error without making use of more advanced debugging tools. For both of these situations, \pkg{SimEngine} includes the following configuration option:

\begin{CodeChunk}
\begin{CodeInput}
R> sim %<>% set_config(stop_at_error=TRUE)
\end{CodeInput}
\end{CodeChunk}

Setting \code{stop_at_error=TRUE} will stop the simulation when it encounters any error. Furthermore, the error will be thrown by \proglang{R} in the usual way, so if the simulation is being run in in RStudio, the built-in debugging tools (such as ``Show Traceback" and ``Rerun with debug") can be used to find and fix the bug. Placing a call to \fct{browser} at the top of the simulation script can also be useful for debugging.


\subsection{Monte Carlo error}\label{sec:mc error}

Statistical simulations are often based on the principle of Monte Carlo approximation; specifically, pseudo-random sampling is used to evaluate the performance of a statistical procedure under a particular data-generating process. The performance of the procedure can be viewed as a statistical parameter and, due to the fact that only a finite number of simulation replicates can be performed, there is uncertainty in any estimate of performance. This uncertainty is often referred to as \textit{Monte Carlo error} (see, e.g., \citealp{lee1999effect}). We can quantify Monte Carlo error using, for example, the standard error of the performance estimator. 

Measuring and reporting Monte Carlo error is a vital component of a simulation study. \pkg{SimEngine} includes an option in the \fct{summary} function to automatically estimate the Monte Carlo standard error for any inferential summary statistic, e.g., estimator bias or confidence interval coverage. The standard error estimates are based on the formulas provided in \citet{morris2019}. If the option \code{mc\_se} is set to \code{TRUE}, estimates of Monte Carlo standard error will be included in the summary data frame, along with associated 95\% confidence intervals based on a normal approximation. 

\begin{CodeChunk}
\begin{CodeInput}
sim <- new_sim()
create_data <- function(n) { rpois(n, lambda=5) }
est_mean <- function(dat) {
  return(mean(dat))
}
sim %<>% set_levels(n=c(10,100,1000))
sim %<>% set_config(num_sim=5)
sim %<>% set_script(function() {
  dat <- create_data(L$n)
  lambda_hat <- est_mean(dat=dat)
  return (list("lambda_hat"=lambda_hat))
})
sim %<>% run()
\end{CodeInput}
\begin{CodeOutput}
  |########################################| 100%
Done. No errors or warnings detected.
\end{CodeOutput}
\begin{CodeInput}
sim %>% summarize(
  list(stat="mse", name="lambda_mse", estimate="lambda_hat", truth=5), 
  mc_se = TRUE
)
\end{CodeInput}
\begin{CodeOutput}
  level_id    n n_reps lambda_mse lambda_mse_mc_se lambda_mse_mc_ci_l
1        1   10      5  0.5020000      0.274178774      -0.0353903966
2        2  100      5  0.0142800      0.012105759      -0.0094472876
3        3 1000      5  0.0031878      0.001919004      -0.0005734471
  lambda_mse_mc_ci_u
1        1.039390397
2        0.038007288
3        0.006949047
\end{CodeOutput}
\end{CodeChunk}

\section*{Computational details}

The results in this paper were obtained using \proglang{R}~4.3.2 with the \pkg{SimEngine}~1.3.0 package. \proglang{R} itself and all packages used are available from the Comprehensive \proglang{R} Archive Network (CRAN) at \url{https://CRAN.R-project.org/}.

\bibliography{refs}

\begin{thebibliography}{17}
\newcommand{\enquote}[1]{``#1''}
\providecommand{\natexlab}[1]{#1}
\providecommand{\url}[1]{\texttt{#1}}
\providecommand{\urlprefix}{URL }
\expandafter\ifx\csname urlstyle\endcsname\relax
  \providecommand{\doi}[1]{doi:\discretionary{}{}{}#1}\else
  \providecommand{\doi}{doi:\discretionary{}{}{}\begingroup \urlstyle{rm}\Url}\fi
\providecommand{\eprint}[2][]{\url{#2}}

\bibitem[{Alfons \emph{et~al.}(2010)Alfons, Templ, and Filzmoser}]{alfons2010object}
Alfons A, Templ M, Filzmoser P (2010).
\newblock \enquote{An object-oriented framework for statistical simulation: The R package simFrame.}
\newblock \emph{Journal of Statistical Software}, \textbf{37}, 1--36.

\bibitem[{Arnold \emph{et~al.}(2011)Arnold, Hogan, Colford, and Hubbard}]{arnold2011simulation}
Arnold BF, Hogan DR, Colford JM, Hubbard AE (2011).
\newblock \enquote{Simulation methods to estimate design power: an overview for applied research.}
\newblock \emph{BMC medical research methodology}, \textbf{11}(1), 1--10.

\bibitem[{Bache and Wickham(2022)}]{bache2022magrittr}
Bache SM, Wickham H (2022).
\newblock \emph{magrittr: A Forward-Pipe Operator for R}.
\newblock R package version 2.0.3, \urlprefix\url{https://CRAN.R-project.org/package=magrittr}.

\bibitem[{Bien(2016)}]{bien2016simulator}
Bien J (2016).
\newblock \enquote{The simulator: an engine to streamline simulations.}
\newblock \emph{arXiv preprint arXiv:1607.00021}.

\bibitem[{Brown and Bye(2023)}]{brown2023simpr}
Brown E, Bye J (2023).
\newblock \emph{simpr: Flexible `Tidyverse'-Friendly Simulations}.
\newblock R package version 0.2.6, \urlprefix\url{https://CRAN.R-project.org/package=simpr}.

\bibitem[{Chalmers and Adkins(2020)}]{chalmers2020writing}
Chalmers RP, Adkins MC (2020).
\newblock \enquote{Writing effective and reliable Monte Carlo simulations with the SimDesign package.}
\newblock \emph{The Quantitative Methods for Psychology}, \textbf{16}(4), 248--280.

\bibitem[{Efron(1979)}]{efron1979}
Efron B (1979).
\newblock \enquote{{Bootstrap Methods: Another Look at the Jackknife}.}
\newblock \emph{The Annals of Statistics}, \textbf{7}(1), 1 -- 26.
\newblock \doi{10.1214/aos/1176344552}.
\newblock \urlprefix\url{https://doi.org/10.1214/aos/1176344552}.

\bibitem[{Hauck and Anderson(1984)}]{hauck1984survey}
Hauck WW, Anderson S (1984).
\newblock \enquote{A survey regarding the reporting of simulation studies.}
\newblock \emph{The American Statistician}, \textbf{38}(3), 214--216.

\bibitem[{Hofert and M{\"a}chler(2013)}]{hofert2013parallel}
Hofert M, M{\"a}chler M (2013).
\newblock \enquote{Parallel and other simulations in R made easy: An end-to-end study.}
\newblock \emph{arXiv preprint arXiv:1309.4402}.

\bibitem[{Koehler \emph{et~al.}(2009)Koehler, Brown, and Haneuse}]{koehler2009assessment}
Koehler E, Brown E, Haneuse SJP (2009).
\newblock \enquote{On the assessment of Monte Carlo error in simulation-based statistical analyses.}
\newblock \emph{The American Statistician}, \textbf{63}(2), 155--162.

\bibitem[{Lee and Young(1999)}]{lee1999effect}
Lee SM, Young GA (1999).
\newblock \enquote{The effect of Monte Carlo approximation on coverage error of double-bootstrap confidence intervals.}
\newblock \emph{Journal of the Royal Statistical Society Series B: Statistical Methodology}, \textbf{61}(2), 353--366.

\bibitem[{Morris \emph{et~al.}(2019)Morris, White, and Crowther}]{morris2019}
Morris TP, White IR, Crowther MJ (2019).
\newblock \enquote{Using simulation studies to evaluate statistical methods.}
\newblock \emph{Statistics in {M}edicine}, \textbf{38}(11), 2074--2102.

\bibitem[{{R Core Team}(2021)}]{Rsoftware}
{R Core Team} (2021).
\newblock \emph{R: A Language and Environment for Statistical Computing}.
\newblock R Foundation for Statistical Computing, Vienna, Austria.
\newblock \urlprefix\url{https://www.R-project.org/}.

\bibitem[{Wakefield(2013)}]{wakefield2013bayesian}
Wakefield J (2013).
\newblock \emph{Bayesian and {F}requentist {R}egression {M}ethods}, volume~23.
\newblock Springer.

\bibitem[{White(1980)}]{white1980heteroskedasticity}
White H (1980).
\newblock \enquote{A heteroskedasticity-consistent covariance matrix estimator and a direct test for heteroskedasticity.}
\newblock \emph{Econometrica: journal of the Econometric Society}, pp. 817--838.

\bibitem[{Zeileis(2004)}]{zeileis2004econometric}
Zeileis A (2004).
\newblock \enquote{Econometric computing with HC and HAC covariance matrix estimators.}
\newblock \emph{Journal of Statistical Software}.

\bibitem[{Zeileis \emph{et~al.}(2020)Zeileis, K{\"o}ll, and Graham}]{zeileis2020various}
Zeileis A, K{\"o}ll S, Graham N (2020).
\newblock \enquote{Various versatile variances: an object-oriented implementation of clustered covariances in R.}
\newblock \emph{Journal of Statistical Software}, \textbf{95}, 1--36.

\end{thebibliography}

\newpage

\begin{appendix}

\section{Simulation-based power calculation}\label{sec:ex1}

Calculating statistical power is a critical step in the design of experiments. For a given study design, the statistical power is defined as the probability that a hypothesis test correctly rejects the null hypothesis (assuming it is false). Sometimes, the sample size for a study is considered fixed, and interest lies in calculating power. More often, investigators want to know what sample size is needed to reject the null hypothesis at a given power level (e.g. 80\% or 90\%). We assume that the reader has some familiarity with statistical hypothesis testing.

For simple study designs (e.g. an individually randomized controlled trial with two groups), formulas exist to calculate the sample size necessary to reject the null hypothesis under certain assumptions on the distribution of the outcome, the effect size, etc. For example, in an experiment comparing means between two groups, the following formula is used to calculate the necessary sample size to reject the null hypothesis of no difference with power $1-\beta$ while maintaining type I error rate $\alpha$, where the outcome variable has means $(\mu_0,\mu_1)$ and variances $(\sigma_0^2,\sigma_1^2)$ in the two groups and $z_q$ denotes the $q$th quantile of the standard normal distribution:

\begin{align}\label{eq1}
    n = \frac{(z_{\alpha/2}+z_\beta)^2(\sigma_0^2+\sigma_1^2)}{(\mu_0-\mu_1)^2}\ .
\end{align}

However, for more complex study designs or analysis plans, sample size formulas may not exist. In these situations, an easier approach is to conduct a simulation-based power calculation. The basic idea is that to repeatedly simulate the entire experiment and calculate the proportion of experiments in which the null hypothesis is rejected; this is the estimated power. Simulating the entire experiment will typically involve generating a dataset and then running an analysis that involves a hypothesis test. Randomness is usually introduced into the process through the dataset generation, although sometimes a population dataset will be fixed and randomness induced by taking samples from that population (e.g., to simulate survey data analyses). Often, the most difficult aspect of a simulation-based power calculation is simulating a dataset that accurately reflects the nuances (e.g., the correlation structure) of the real dataset.

To calculate sample size at a fixed power level (e.g., 90\%), a ``guess and check'' approach may be used. With this approach, the simulation is run with an arbitrarily-selected sample size $n_1$ to estimate the power. If power is estimated to be lower than 90\%, a new, larger value $n_2$ is selected and the simulation is run again again. This procedure is repeated until the estimated power is roughly 90\%.

We illustrate this process by simulating a randomized controlled trial and comparing the results to what we would have attained using (\ref{eq1}). First, we declare a new simulation object and write a function to generate data:

\begin{CodeChunk}
\begin{CodeInput}
sim <- new_sim()

create_rct_data <- function(n, mu_0, mu_1, sigma_0, sigma_1) {
  group <- sample(rep(c(0,1),n))
  outcome <- (1-group) * rnorm(n=n, mean=mu_0, sd=sigma_0) +
             group * rnorm(n=n, mean=mu_1, sd=sigma_1)
  return(data.frame("group"=group, "outcome"=outcome))
}

create_rct_data(n=3, mu_0=3, mu_1=4, sigma_0=0.1, sigma_1=0.1)
\end{CodeInput}
\begin{CodeOutput}
  group  outcome
1     1 4.073832
2     0 2.917953
3     1 3.969461
4     0 3.032951
5     0 2.917953
6     1 3.969461
\end{CodeOutput}
\end{CodeChunk}

Next, we add a function that takes a dataset generated by the \fct{create\_rct\_data} function and runs a statistical test to determine whether to reject the null hypothesis:

\begin{CodeChunk}
\begin{CodeInput}
run_test <- function(data) {
  test_result <- t.test(outcome~group, data=data)
  return(as.integer(test_result$p.value<0.05))
} 
\end{CodeInput}
\end{CodeChunk}

Next, we write the simulation script and tell \pkg{SimEngine} to run 1,000 simulation replicates each for four sample sizes.

\begin{CodeChunk}
\begin{CodeInput}
sim %<>% set_script(function() {
  data <- create_rct_data(n=L$n, mu_0=17, mu_1=18, sigma_0=2, sigma_1=2)
  reject <- run_test(data)
  return (list("reject"=reject))
})
sim %<>% set_levels(n=c(20,40,60,80))
sim %<>% set_config(num_sim=1000)
\end{CodeInput}
\end{CodeChunk}

Next, we run the simulation. After obtaining results, we calculate power by averaging the \code{reject} variable using the \fct{summarize} function, which indicates the percentage of simulations in which the null hypothesis was rejected.

\begin{CodeChunk}
\begin{CodeInput}
sim %<>% run()
\end{CodeInput}
\begin{CodeOutput}
  |########################################| 100%
Done. No errors or warnings detected.
\end{CodeOutput}
\begin{CodeInput}
power_sim <- sim %>% summarize(
  list(stat="mean", name="power", x="reject")
)
print(power_sim)
\end{CodeInput}
\begin{CodeOutput}
  level_id  n n_reps power
1        1 20   1000 0.364
2        2 40   1000 0.578
3        3 60   1000 0.736
4        4 80   1000 0.844
\end{CodeOutput}
\end{CodeChunk}

We can compare the results to what we obtain from (\ref{eq1}).

\begin{CodeChunk}
\begin{CodeInput}
power_formula <- sapply(c(20,40,60,80), function(n) {
  pnorm(sqrt((n*(17-18)^2)/(2^2+2^2)) - qnorm(0.025, lower.tail=F))
})
library(ggplot2)
ggplot(data.frame(
  n = rep(c(20,40,60,80), 2),
  power = c(power_sim$power, power_formula),
  which = rep(c("Simulation","Formula"), each=4)
), aes(x=n, y=power, color=factor(which))) +
  geom_line() +
  theme_bw() +
  labs(color="Method", y="Power", x="Sample size (per group)")
\end{CodeInput}
\end{CodeChunk}

\begin{figure}[ht!]
    \centering
    \includegraphics[width=0.75\linewidth]{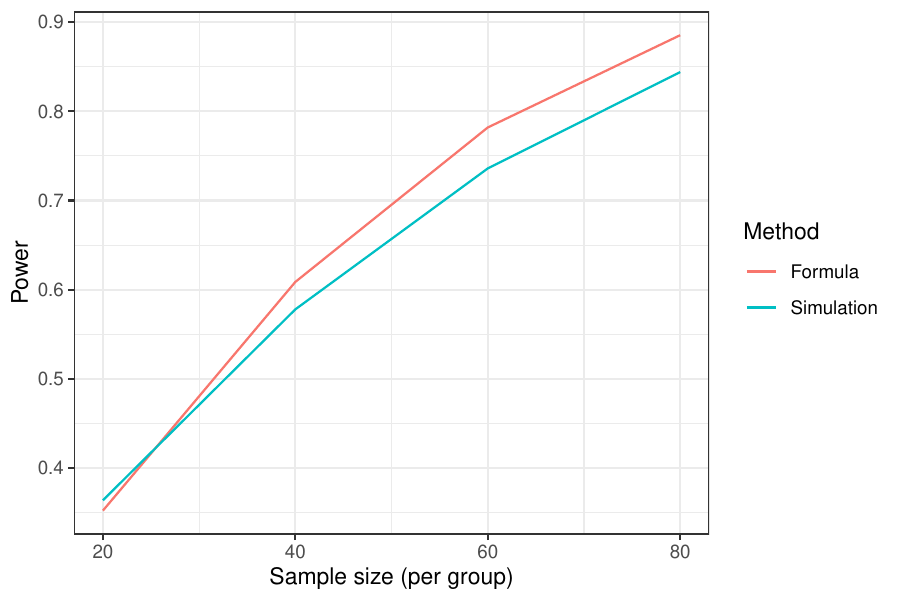}
    \label{fig_ex1_powercalc}
\end{figure}

Of course, real applications will typically involve much more complex data generating mechanisms and analyses, but the same basic principles illustrated in the code above will generally apply to simulation-based power calculations.

\section{Comparing two standard error estimators}\label{sec:ex2}

When developing a novel statistical method, we often wish to compare our proposed method with one or more existing methods. This serves to highlight the differences between our method and whatever is used in common practice. Generally, we wish to examine realistic settings, motivated by statistical theory, in which the novel method confers some advantage over the alternatives. 

In this example, we will consider the problem of estimating the variance-covariance matrix of the least-squares estimator in linear regression. We assume the reader has some familiarity with linear regression. 

Suppose the dataset consists of $n$ independent observations $\{(Y_1, X_1), (Y_2, X_2), \ldots, (Y_n, X_n)\}$, where $X$ and $Y$ are both scalar variables. A general linear regression model posits that
$$Y_i = \beta_0 + \beta_1 X_i + \epsilon_i\ ,$$
where $\epsilon_i$ is a mean-zero noise term with variance $\sigma^2_i$. We refer to this as a heteroskedastic model, since the variances need not be equal across all $i$. This is the true data-generating model. We note that a more restrictive (but misspecified) model assumes that there is a common variance $\sigma^2$ such that $\sigma^2_i = \sigma^2$ for all $i$. We refer to this incorrect model as the homoskedastic model. 

In linear regression, the least-squares method is often used to construct an estimate $\hat{\beta}$ of the coefficient vector $\beta$. For the purposes of building confidence intervals and performing hypothesis tests, we may also wish to estimate the standard error of the least squares estimator. Perhaps the two most common ways to do this are: 
\begin{enumerate}
    \item Use a model-based standard error that assume homoskedasticity. This is the estimator used by default in most statistical software, including the \fct{lm} function in \proglang{R}. 
    \item Use a so-called sandwich standard error \citep{white1980heteroskedasticity}. Statistical theory shows that this estimator will be consistent even under heteroskedasticity.
\end{enumerate}
We will carry out a small simulation study to compare these two estimators. 

We start by declaring a new simulation object and writing a function that generates some data according to the heteroskedastic model. In this simulation, $\sigma^2_i$ is larger for larger values of $X_i$. 

\begin{CodeChunk}
\begin{CodeInput}
sim <- new_sim()

create_regression_data <- function(n) {
  beta <- c(-1, 10)
  x <- rnorm(n)
  sigma2 <- exp(x)
  y <- rnorm(n=n, mean=(beta[1]+beta[2]*x), sd = sqrt(sigma2))
  return(data.frame(x=x, y=y))
}
\end{CodeInput}
\end{CodeChunk}

For a sense of what this sort of heteroskedasticity looks like empirically, we generate a dataset, fit a linear regression model, and make a scatterplot of the residuals against $X$.

\begin{CodeChunk}
\begin{CodeInput}
dat <- create_regression_data(n=500)
linear_model <- lm(y~x, data=dat)
dat$residuals <- linear_model$residuals
library(ggplot2)
ggplot(dat, aes(x=x, y=residuals)) +
  geom_point() +
  theme_bw() +
  labs(x="x", y="residual")
\end{CodeInput}
\end{CodeChunk}

\begin{figure}[ht!]
    \centering
    \includegraphics[width=0.75\linewidth]{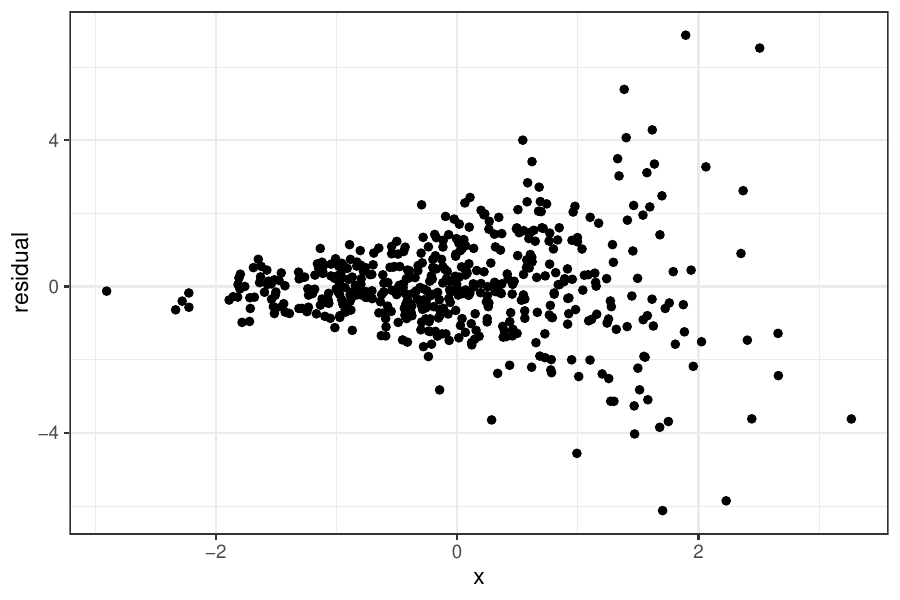}
    \label{fig_ex2_scatterplot}
\end{figure}

Now, we declare two methods: one returns the least squares estimate $\hat{\beta}$ and model-based estimate of the variance-covariance matrix of $\hat{\beta}$, and the other returns $\hat{\beta}$ and the sandwich estimate from \fct{vcovHC} in the \pkg{sandwich} package \citep{zeileis2004econometric, zeileis2020various}. 

\begin{CodeChunk}
\begin{CodeInput}
model_vcov <- function(data) {
  mod <- lm(y~x, data=data)
  return(list("coef"=mod$coefficients, "vcov"=diag(vcov(mod))))
}
sandwich_vcov <- function(data) {
  mod <- lm(y~x, data=data)
  return(list("coef"=mod$coefficients, "vcov"=diag(vcovHC(mod))))
}
\end{CodeInput}
\end{CodeChunk}

Next, we write the simulation script. This script returns a point estimate and a standard error estimate for both the intercept parameter $\beta_0$ and the slope parameter $\beta_1$. We tell \pkg{SimEngine} to run 500 simulation replicates for each of four sample sizes. It is important to use the \code{seed} argument in \fct{set\_config} so that our results will be reproducible. In addition, we use the \code{packages} option to load the \pkg{sandwich} package. Loading packages via \fct{set\_config} (as opposed to running \code{library(sandwich)}) is required if running simulations in parallel. Finally, we run the simulation.  

\begin{CodeChunk}
\begin{CodeInput}
sim %<>% set_script(function() {
  data <- create_regression_data(n=L$n)
  estimates <- use_method(L$estimator, list(data))
  return(list(
    "beta0_est" = estimates$coef[1],
    "beta1_est" = estimates$coef[2],
    "beta0_se_est" = sqrt(estimates$vcov[1]),
    "beta1_se_est" = sqrt(estimates$vcov[2])
  ))
})
sim %<>% set_levels(
  estimator = c("model_vcov", "sandwich_vcov"),
  n = c(50, 100, 500, 1000)
)
sim %<>% set_config(
  num_sim = 500,
  seed = 24,
  packages = c("sandwich")
)
sim %<>% run()
\end{CodeInput}
\begin{CodeOutput}
  |########################################| 100%
Done. No errors or warnings detected.
\end{CodeOutput}
\end{CodeChunk}

Now we can summarize the results using \fct{summarize}. There are two main quantities of interest. The primary purpose of the standard error estimate for $\hat{\beta}$ is to form confidence intervals, so we examine (1) the average half-width of the resulting interval (simply 1.96 times the average standard error estimate across simulation replicates), and (2) the estimated coverage of the interval, which is simply the proportion of simulation replicates in which the interval contains the true value of $\beta$. We focus on 95\% confidence intervals in this simulation. 

\begin{CodeChunk}
\begin{CodeInput}
summarized_results <- sim %>% summarize(
  list(stat="mean", name="mean_se_beta0", x="beta0_se_est"),
  list(stat="mean", name="mean_se_beta1", x="beta1_se_est"),
  list(stat="coverage", name="cov_beta0", estimate="beta0_est",
       se="beta0_se_est", truth=-1),
  list(stat="coverage", name="cov_beta1", estimate="beta1_est",
       se="beta1_se_est", truth=10)
)

print(summarized_results)
\end{CodeInput}
\begin{CodeOutput}
  level_id     estimator    n n_reps mean_se_beta0 mean_se_beta1 cov_beta0
1        1    model_vcov   50    500    0.18100830    0.18233202     0.946
2        2 sandwich_vcov   50    500    0.18235693    0.24320294     0.938
3        3    model_vcov  100    500    0.12724902    0.12778192     0.946
4        4 sandwich_vcov  100    500    0.12805679    0.17341936     0.942
5        5    model_vcov  500    500    0.05704424    0.05697317     0.946
6        6 sandwich_vcov  500    500    0.05748439    0.07960900     0.948
7        7    model_vcov 1000    500    0.04055718    0.04059151     0.960
8        8 sandwich_vcov 1000    500    0.04052126    0.05721446     0.958
  cov_beta1
1     0.848
2     0.922
3     0.862
4     0.946
5     0.836
6     0.940
7     0.860
8     0.958
\end{CodeOutput}
\end{CodeChunk}

To visualize the results, we set up a plotting function.

\begin{CodeChunk}
\begin{CodeInput}
library(dplyr)
library(tidyr)
plot_results <- function(summarized_results, which_graph, n_est) {
  if (n_est == 3) {
    values <- c("#999999", "#E69F00", "#56B4E9")
    breaks <- c("model_vcov", "sandwich_vcov", "bootstrap_vcov")
    labels <- c("Model-based", "Sandwich", "Bootstrap")
  } else {
    values <- c("#999999", "#E69F00")
    breaks <- c("model_vcov", "sandwich_vcov")
    labels <- c("Model-based", "Sandwich")
  }
  if (which_graph == "width") {
    summarized_results %>%
    pivot_longer(
      cols = c("mean_se_beta0", "mean_se_beta1"),
      names_to = "parameter",
      names_prefix = "mean_se_"
    ) %>%
    mutate(value_j = jitter(value, amount = 0.01)) %>%
    ggplot(aes(x=n, y=1.96*value_j, color=estimator)) +
    geom_line(aes(linetype=parameter)) +
    geom_point() +
    theme_bw() +
    ylab("Average CI width") +
    xlab("Sample size") + 
    scale_color_manual(
      values = values,
      breaks = breaks,
      name = "SE estimator",
      labels = labels
    ) +
    scale_linetype_discrete(
      breaks = c("beta0", "beta1"),
      name = "Parameter",
      labels = c(expression(beta[0]), expression(beta[1]))
    )
  } else {
    summarized_results %>%
    pivot_longer(
      cols = c("cov_beta0","cov_beta1"),
      names_to = "parameter",
      names_prefix = "cov_"
    ) %>%
    mutate(value_j = jitter(value, amount = 0.01)) %>%
    ggplot(aes(x=n, y=value, color=estimator)) +
    geom_line(aes(linetype = parameter)) +
    geom_point() +
    theme_bw() +
    ylab("Coverage") +
    xlab("Sample size") +
    scale_color_manual(
      values = values,
      breaks = breaks,
      name = "SE estimator",
      labels = labels
    ) +
    scale_linetype_discrete(
      breaks = c("beta0", "beta1"),
      name = "Parameter",
      labels = c(expression(beta[0]), expression(beta[1]))
    ) +
    geom_hline(yintercept=0.95)
  }
}
\end{CodeInput}
\end{CodeChunk}

We then use the plotting function to make figures. 

\begin{CodeChunk}
\begin{CodeInput}
plot_results(summarized_results, "width", 2)
\end{CodeInput}
\end{CodeChunk}

\begin{figure}[ht!]
    \centering
    \includegraphics[width=0.75\linewidth]{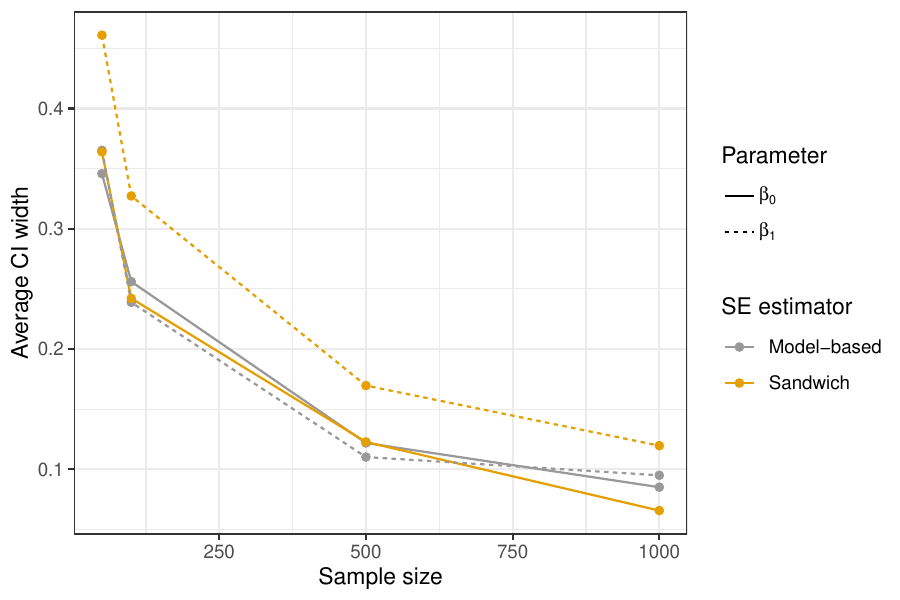}
    \label{fig_ex2_ci_width}
\end{figure}

\begin{CodeChunk}
\begin{CodeInput}
plot_results(summarized_results, "coverage", 2)
\end{CodeInput}
\end{CodeChunk}

\begin{figure}[ht!]
    \centering
    \includegraphics[width=0.75\linewidth]{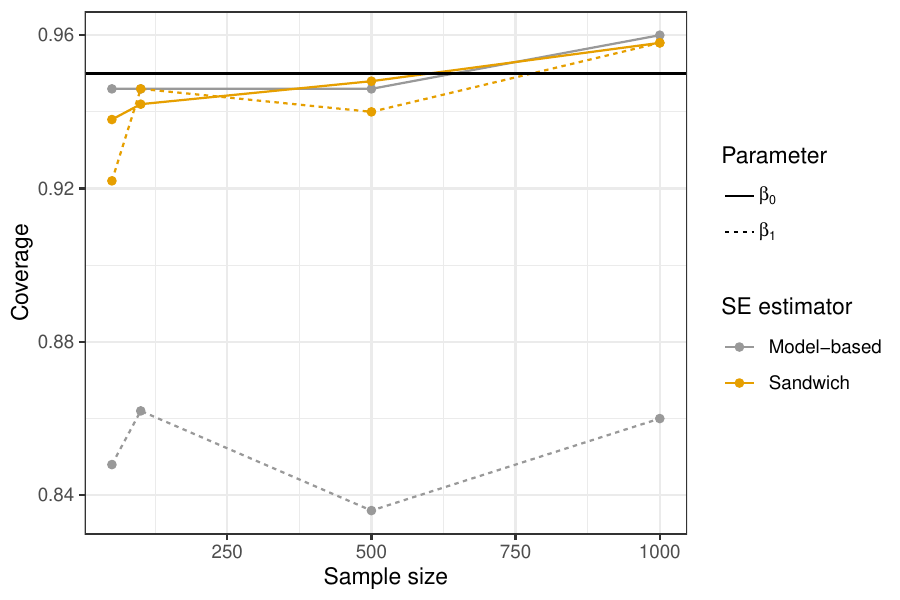}
    \label{fig_ex2_ci_coverage}
\end{figure}

Looking at these plots, we see that the sandwich method results in a wider interval, on average, for $\beta_1$. In terms of coverage, the sandwich estimator achieves near nominal coverage for both parameters, while there is moderate undercoverage for $\beta_1$ using the model-based estimator. 

The bootstrap is another popular approach to estimating standard errors \citep{efron1979}. We can add a bootstrap method and use \fct{update\_sim} to run the new simulation replicates without re-running any of the previous work. We do this by including the new estimator in the simulation levels. Since the bootstrap can be computationally intensive, we use parallelization. This requires specifying the option `\code{parallel = TRUE} in \fct{set\_config}.

\begin{CodeChunk}
\begin{CodeInput}
bootstrap_vcov <- function(data) {
  mod <- lm(y~x, data=data)
  boot_ests <- matrix(NA, nrow=100, ncol=2)
  for (j in 1:100) {
    indices <- sample(1:nrow(data), size=nrow(data), replace=TRUE)
    boot_dat <- data[indices,]
    boot_mod <- lm(y~x, data=boot_dat)
    boot_ests[j,] <- boot_mod$coefficients
  }
  boot_v1 <- var(boot_ests[,1])
  boot_v2 <- var(boot_ests[,2])
  return(list("coef"=mod$coefficients, "vcov"=c(boot_v1, boot_v2)))
}
sim %<>% set_levels(
  estimator = c("model_vcov", "sandwich_vcov", "bootstrap_vcov"),
  n = c(50, 100, 500, 1000)
)
sim %<>% set_config(
  num_sim = 500,
  seed = 24,
  parallel = TRUE,
  n_cores = 2,
  packages = c("sandwich")
)
sim %<>% update_sim()
\end{CodeInput}
\begin{CodeOutput}
Done. No errors or warnings detected.
\end{CodeOutput}
\end{CodeChunk}

Now that the bootstrap results are included in the simulation object, we can look at the updated results. 

\begin{CodeChunk}
\begin{CodeInput}
summarized_results <- sim %>% summarize(
  list(stat="mean", name="mean_se_beta0", x="beta0_se_est"),
  list(stat="mean", name="mean_se_beta1", x="beta1_se_est"),
  list(stat="coverage", name="cov_beta0", estimate="beta0_est",
       se="beta0_se_est", truth=-1),
  list(stat="coverage", name="cov_beta1", estimate="beta1_est",
       se="beta1_se_est", truth=10)
)

plot_results(summarized_results, "width", 3)
\end{CodeInput}
\end{CodeChunk}

\begin{figure}[ht!]
    \centering
    \includegraphics[width=0.75\linewidth]{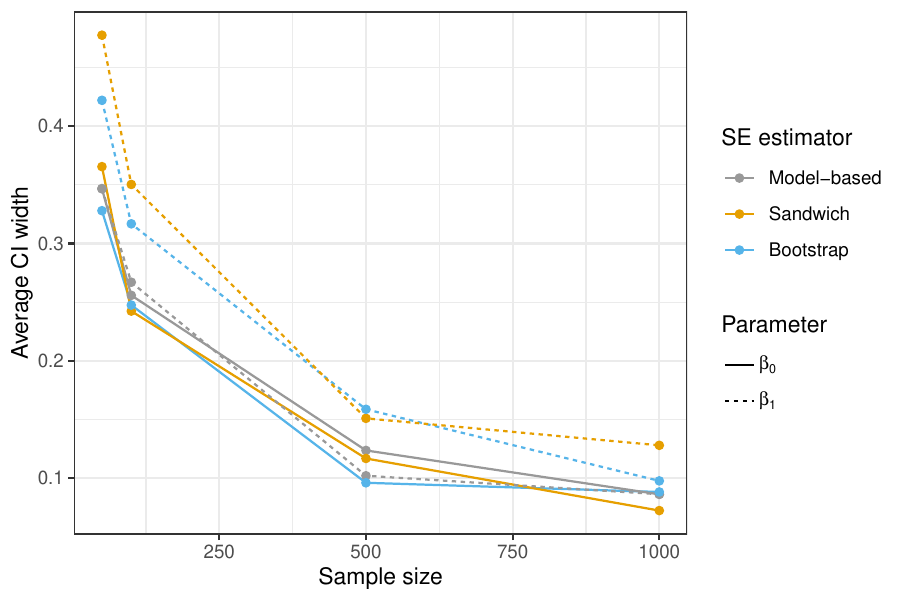}
    \label{fig_ex2_ci_width_2}
\end{figure}

\begin{CodeChunk}
\begin{CodeInput}
plot_results(summarized_results, "coverage", 3)
\end{CodeInput}
\end{CodeChunk}

\begin{figure}[ht!]
    \centering
    \includegraphics[width=0.75\linewidth]{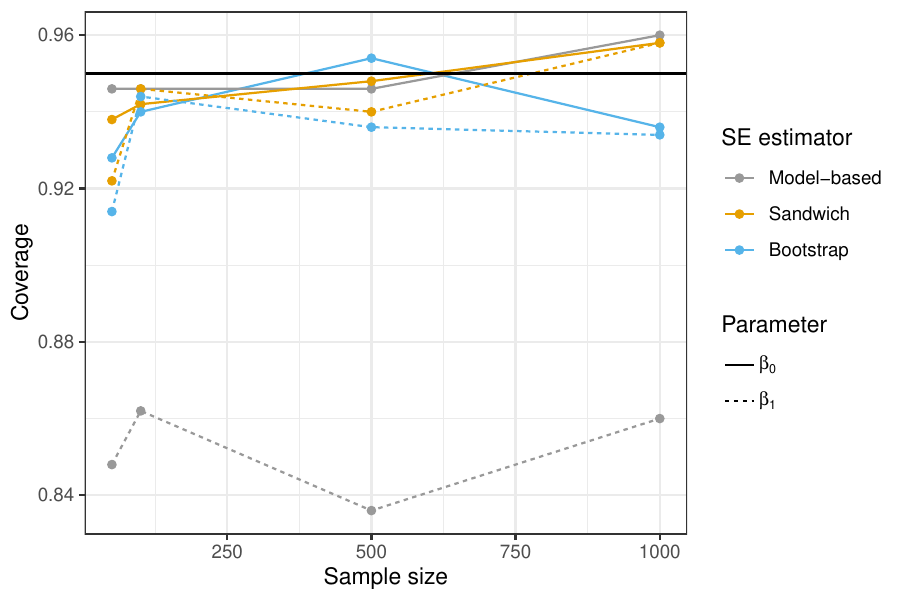}
    \label{fig_ex2_ci_coverage_2}
\end{figure}

Like the sandwich estimator, the bootstrap results in wider intervals for $\beta_1$, but is much closer to achieving 95\% coverage compared to the model-based estimator. 

\end{appendix}

\end{document}